\documentclass[preprint,aps,showpacs,prd,nofootinbib]{revtex4}

\usepackage[latin1]{inputenc}
\usepackage{bbm}

\newcommand{\be}{\begin{eqnarray}} 
\newcommand{\ee}{\end{eqnarray}}

\newcommand{\bmp}{\noindent\begin{minipage}{16cm}}
\newcommand{\emp}{\end{minipage}\vskip 7mm} 

\newcommand{\titel}[1]{}

\usepackage{graphicx}
\usepackage{subfigure}

\begin{document}


\title{
Light composite Higgs and precision electroweak measurements on the
Z resonance: An update
}

\author{Dennis D. {\sc Dietrich}}
\author{Francesco {\sc Sannino}}
\affiliation{The Niels Bohr Institute, Copenhagen, Denmark}
\author{Kimmo {\sc Tuominen}}
\affiliation{Department of Physics, University of Jyv\"askyl\"a, Finland}
\affiliation{Helsinki Institute of Physics, University of Helsinki, Finland}

\date{October 17, 2005}


\begin{abstract}

We update our analysis of technicolour theories with techniquarks in higher
dimensional representations of the technicolour gauge group in the light of
the new electroweak precision data on the Z resonance.


\end{abstract}


\maketitle


\section{Introduction}

In \cite{Dietrich:2005jn}, we analysed technicolour theories
\cite{TC,TCrev} for the breaking of the electroweak symmetry with the
techniquarks in higher representations of the gauge group \cite{higher}. We
identified theories with two techniflavours in the two-index symmetric
representation of SU$_\mathrm{T}$(2) as being consistent with the electroweak
precision data available to that date \cite{Eidelman:2004wy}. At the same
time, this theory is quasi-conformal \cite{Sannino:2004qp,Hong:2004td} (walking
coupling). This feature is a necessity for being able to generate
sufficiently high masses for the ordinary fermions. It also helps avoiding inconsistently large
flavour-changing neutral currents and lepton number violation
due to extended technicolour interactions \cite{Cohen:1988sq,etc}. Remarkably, also
due to the walking, this special choice for the number of technicolours,
techniflavours, and the representation leads to a predicted mass of the
(non-elementary) Higgs of only 150GeV
\footnote{It is relevant to note that even for technicolour theories resembling
QCD the scalar sector is not simply described by just a heavy composite
Higgs. One might also observe for these type of technicolour theories at
CERN-LHC a scalar substantially lighter than one TeV. This composite scalar
is the direct analog of the QCD scalar $f_0(600)$ \cite{sigma} and it
is expected to be a four quark object.}.
For this particular set-up, in order
to avoid the Witten anomaly \cite{Witten:fp}, an additional family of leptons
has to be included, which, amongst other things, provides possible
non-hadronic components of
dark matter. For the masses of these leptons we were able to make
accurate predictions based on the electroweak precision data
at hand. Since
then new data has become available \cite{unknown:2005em}. It, at the 68$\%$
level of confidence, leads to a considerably larger parameter space for the
lepton masses than was expected previously at the 90$\%$ level of confidence.

Widely independently of this, in \cite{Dietrich:2005jn} we had
given an overview of the expected spectrum of technicolour-neutral particles.
However, there, we did not mention that any number of techniquarks in the
two-index symmetric representation of SU$_\mathrm{T}$(2) can be made
technicolour neutral by adding technigluons. This is so since for 
SU$_\mathrm{T}$(2) the two-index symmetric representation 
coincides with the adjoint
representation. The potentially lowest-lying technihadrons of this kind are
bound states made out of one techniquark and technigluons. 
From the viewpoint of the standard model such bound states possess only
weak interactions and mimic an additional lepton family. However, they also  
interact directly via the technicolour sector.


\section{Analysis for the new data}

After having fixed the number of particles, the gauge group, and the
representation, it still remains to define the hypercharge assignment 
which is constrained but not fixed entirely
by imposing the absence of gauge anomalies. 
We have studied the following cases
\cite{Dietrich:2005jn,Sannino:2005dy}: (I) a standard-model like case, in which the leptons
are neutral and singly negatively charged, respectively; (II) a case, in
which the leptons carry half elementary charges with opposite signs; (III) a
singly and a doubly negatively charged lepton. Apart from various hadronic
objects in all cases, in (I) the fourth neutrino is a natural dark matter
candidate.

\begin{figure*}[h]
\centering
 \subfigure[~Perturbative]{
  \includegraphics[width=7cm]{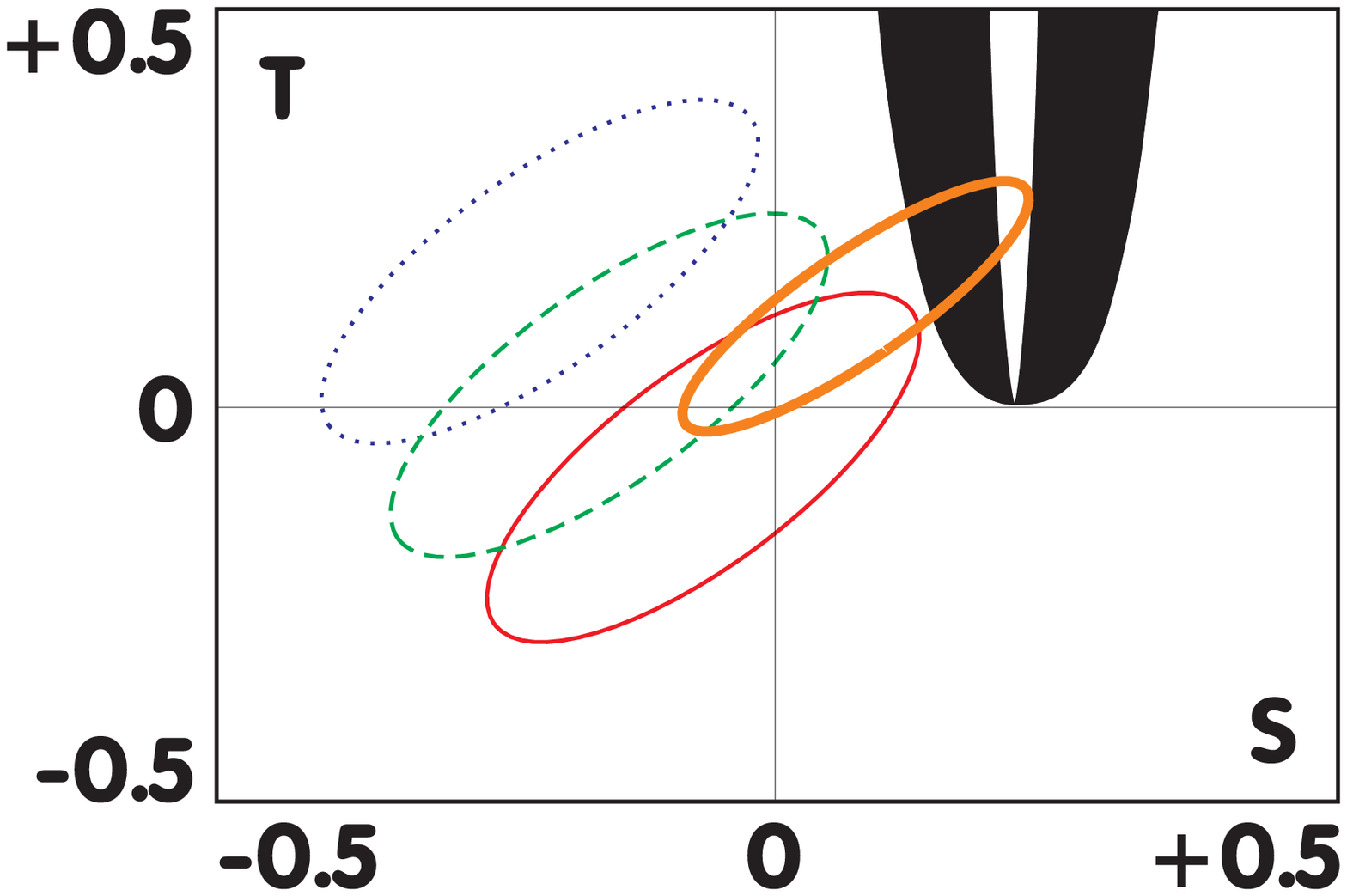}
 }
 \qquad
 \subfigure[~Non-perturbative]{
  \includegraphics[width=7cm]{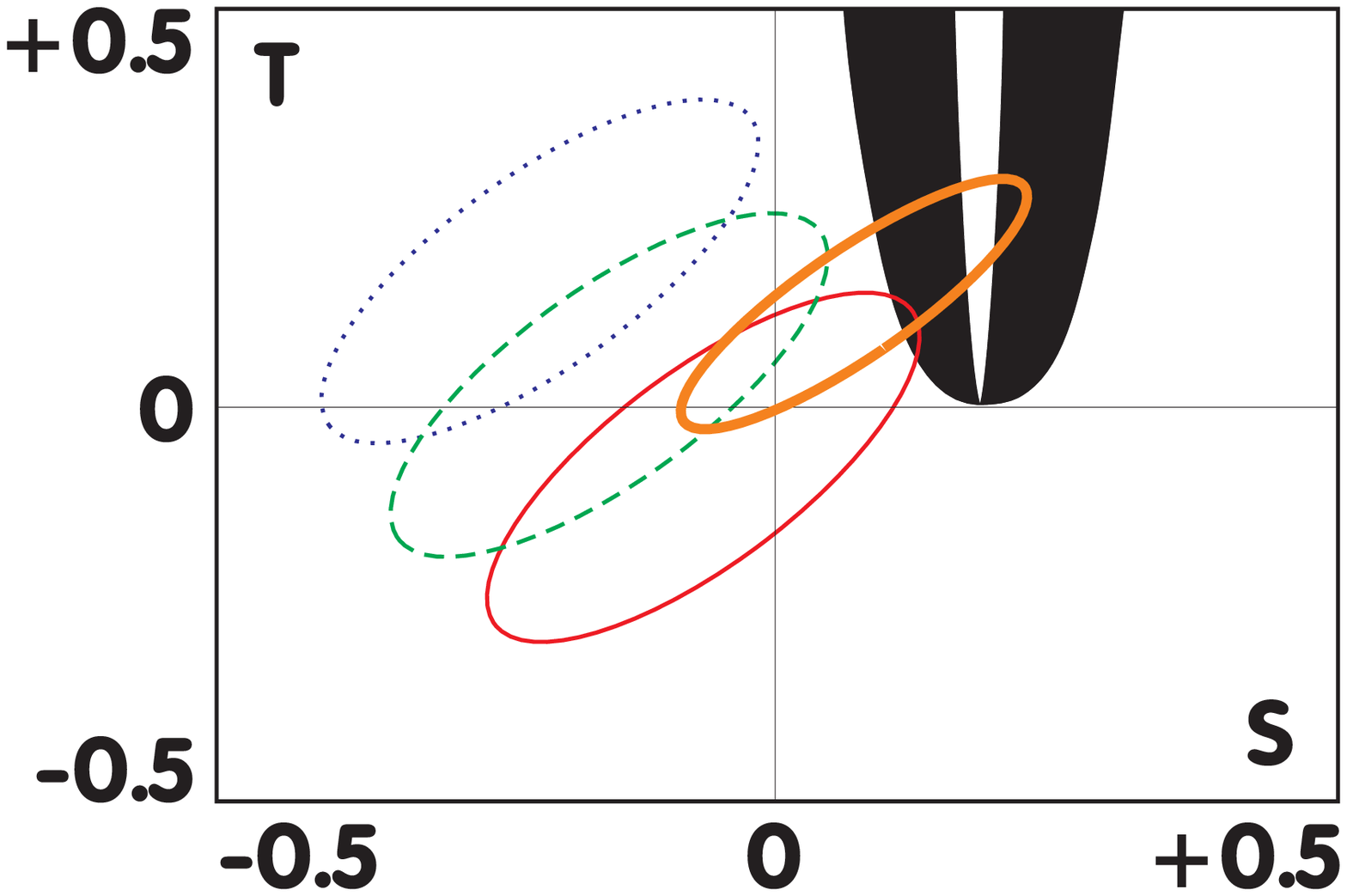}
 }
\caption{\underline{Standard-model-like charge assignment.} {\it Left Panel}:
The area shaded in black corresponds to the accessible range for $S$
and $T$ with the masses of the extra neutrino and extra electron taken from
$m_Z$ to $10 m_Z$. The perturbative estimate for the contribution to $S$ from
techniquarks equals $1/2\pi$. The three staggered ellipses are the $90$\%
confidence level contours for the former global fit to the electroweak
precision data \cite{Eidelman:2004wy} with $U$ kept at $0$. The values of $U$
in our model lie typically between $0$ and $0.05$ whence they are consistent
with these contours. These contours from bottom to top are for Higgs masses of
$m_H = 117$, $340$, $1000$ GeV, respectively. The smaller ellipse to the
upper right is the 68\% confidence level contour for the new global fit to
electroweak precision data \cite{unknown:2005em} with $U=0$ and for a Higgs
$m_H=150$ GeV as predicted for our model.
{\it Right Panel}: With non-perturbative corrections to the $S$ parameter taken
into account in the technicolour sector of the theory.}
\label{smleptons}
\end{figure*}

\begin{figure*}
\centering
 \subfigure[~Perturbative]{
  \includegraphics[width=7cm]{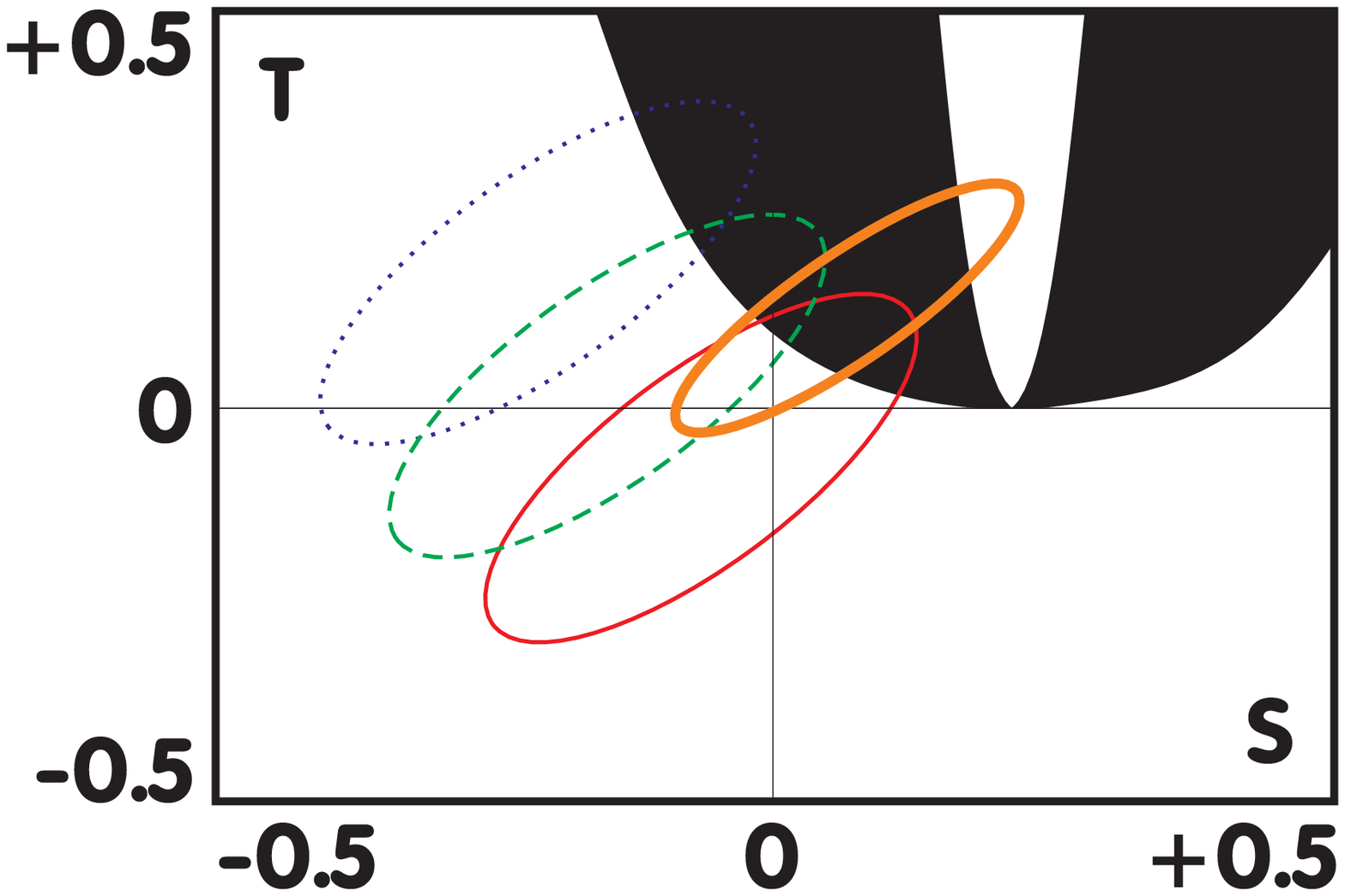}
 }
 \qquad
 \subfigure[~Non-perturbative]{
  \includegraphics[width=7cm]{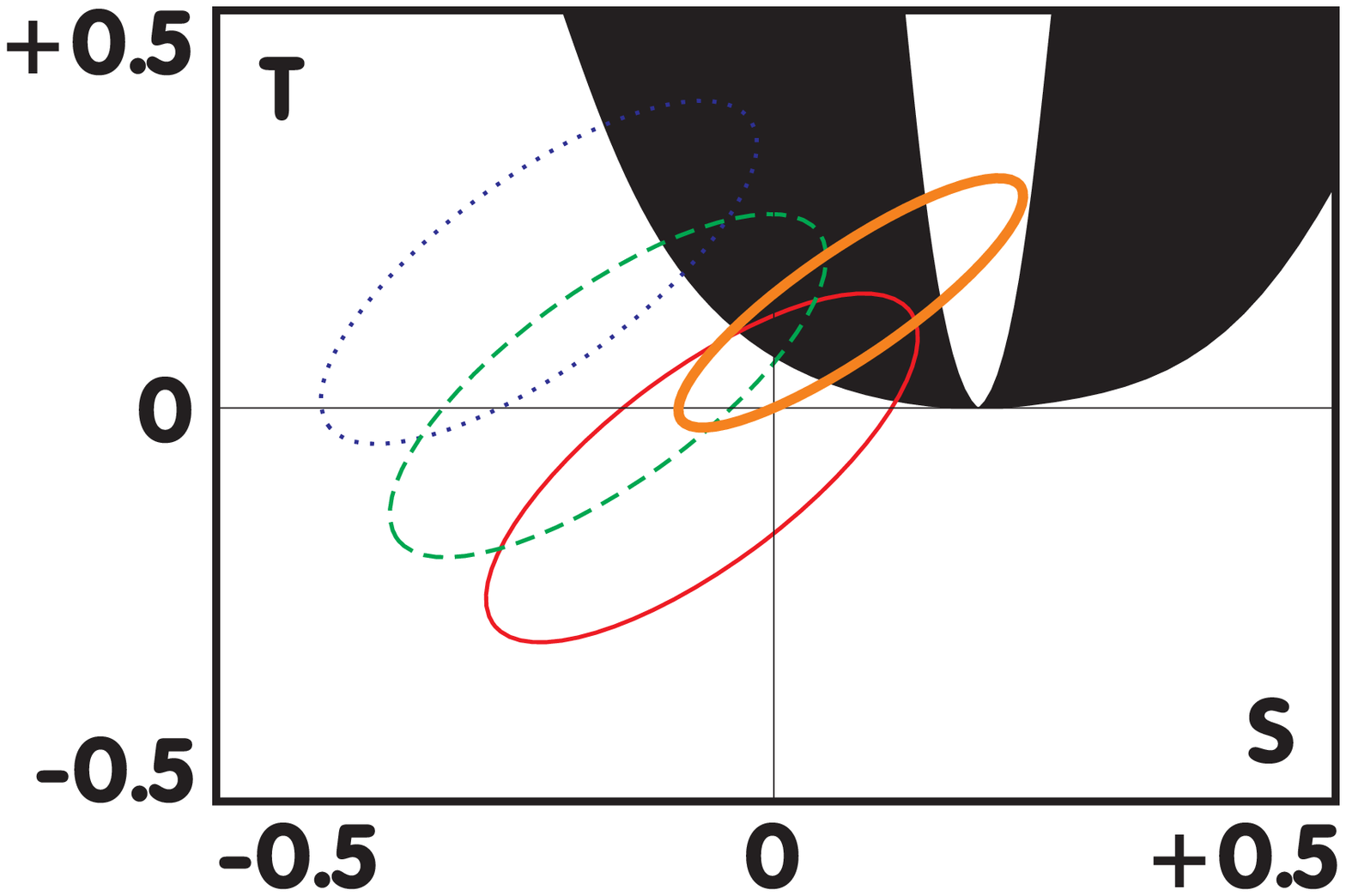}
 }
\caption{\underline{Leptons with integer charges.} {\it Left Panel}: The
parabolic area shaded in black corresponds to the accessible range for $S$
and $T$ with the masses of the extra neutrino and extra electron taken from
$m_Z$ to $10 m_Z$. The perturbative estimate for the contribution to $S$ from
techniquarks equals $1/2\pi$. The three staggered ellipses are the $90$\%
confidence level contours for the former global fit to the electroweak
precision data \cite{Eidelman:2004wy} with $U$ kept at $0$. The values of $U$
in our model lie typically between $0$ and $0.05$ whence they are consistent
with these contours. These contours from bottom to top are for Higgs masses
of $m_H = 117$, $340$, $1000$ GeV, respectively. The smaller ellipse to the
upper right is the 68\% confidence level contour for the new global fit to
electroweak precision data \cite{unknown:2005em} with $U=0$ and for a Higgs
$m_H=150$GeV as predicted for our model.
{\it Right Panel}: With non-perturbative corrections to the $S$ parameter taken
into account in the technicolour sector of the theory.
}
\label{frleptons}
\end{figure*}

The black shaded areas in Figs.~\ref{smleptons} and \ref{frleptons} show
the accessible range of values of the oblique parameters $S$ and $T$
\cite{Peskin:1990zt} \footnote{These parameters measure the contribution of
the non-standard-model particles to the vacuum polarisation of the gauge
bosons. Roughly speaking, $S$ is connected to the mixing of the photon with
the Z-boson and $T$ to contributions to the violation of the isospin
symmetry.} for degenerate techniquarks and if the masses of the leptons are
varied independently in the range from one to ten Z-boson masses. The value
of the third oblique parameter $U$ is close to zero for our model,
consistent with presented data. The larger
staggered ellipses in all of these plots are the 90$\%$ confidence level
contours from the global fit to the data presented in \cite{Eidelman:2004wy}.
The smaller single ellipse represents the 68$\%$ confidence level contour
from the new global fit in \cite{unknown:2005em}.

\begin{figure*}
 \centering
 \subfigure[~Standard-model-like (I)]{
  \includegraphics[width=5cm]{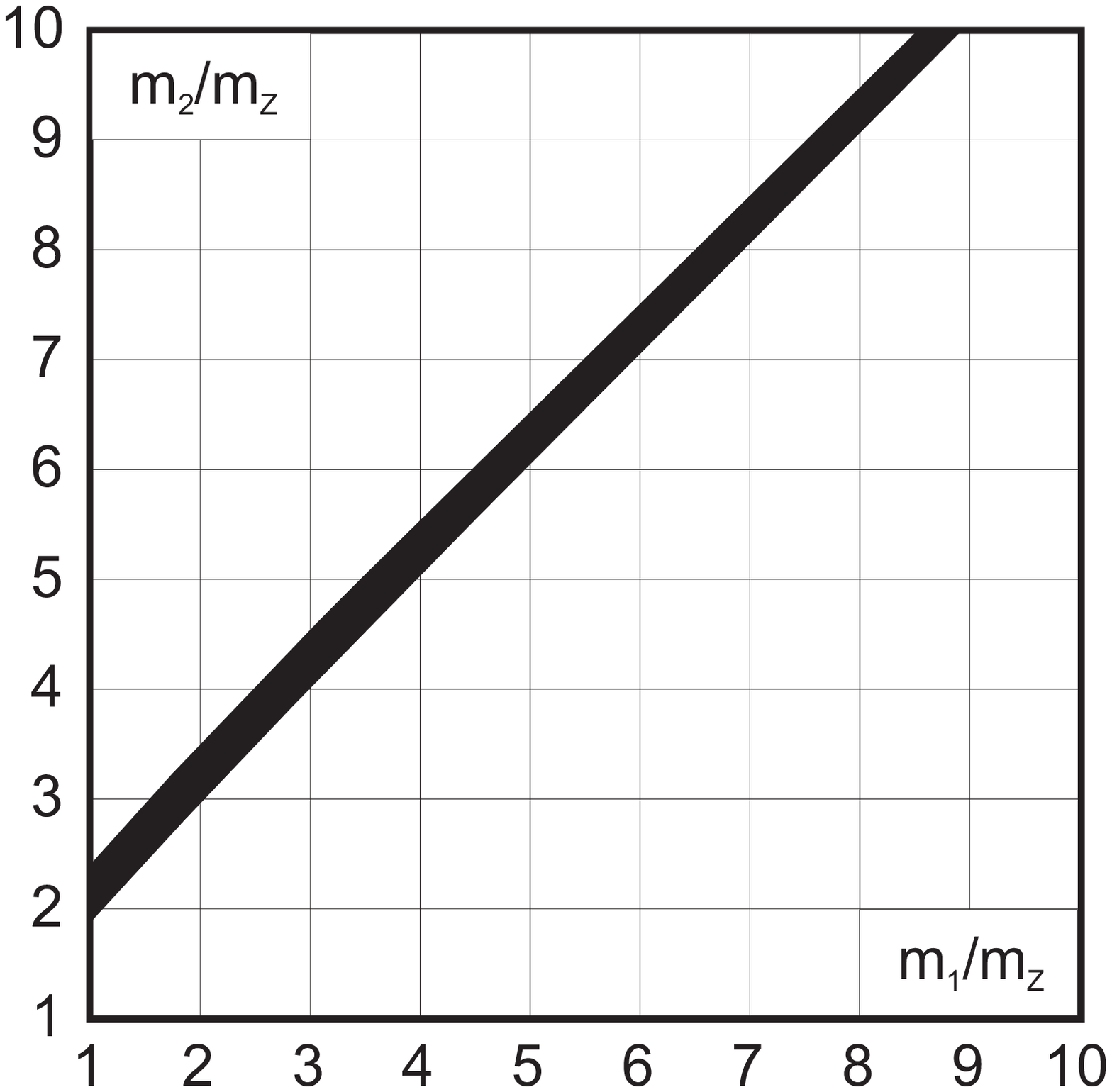}
 }
 \subfigure[~Integer charges (III)]{
  \includegraphics[width=5cm]{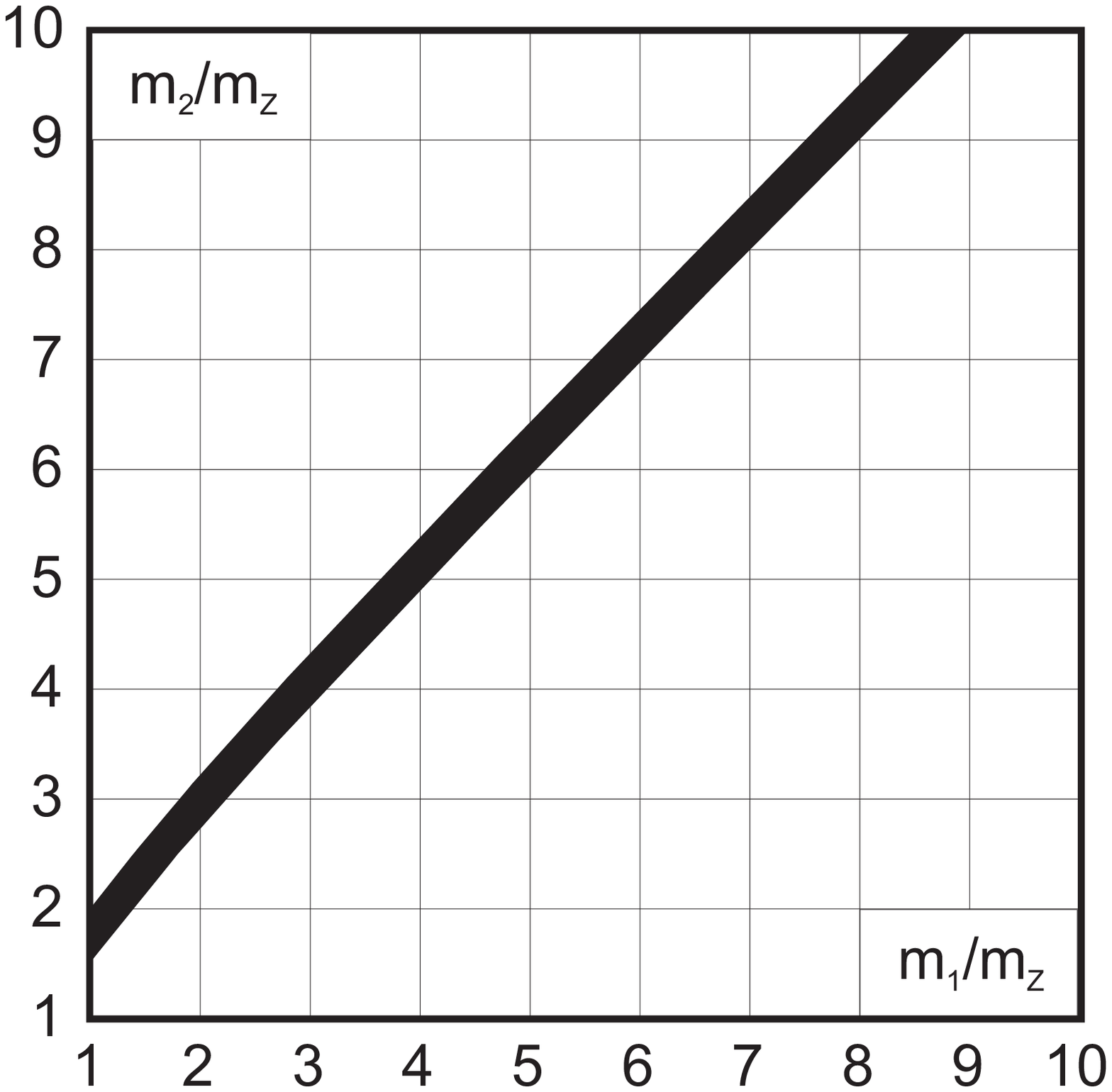}
 }
 \subfigure[~Fractional charges (II)]{
  \includegraphics[width=5cm]{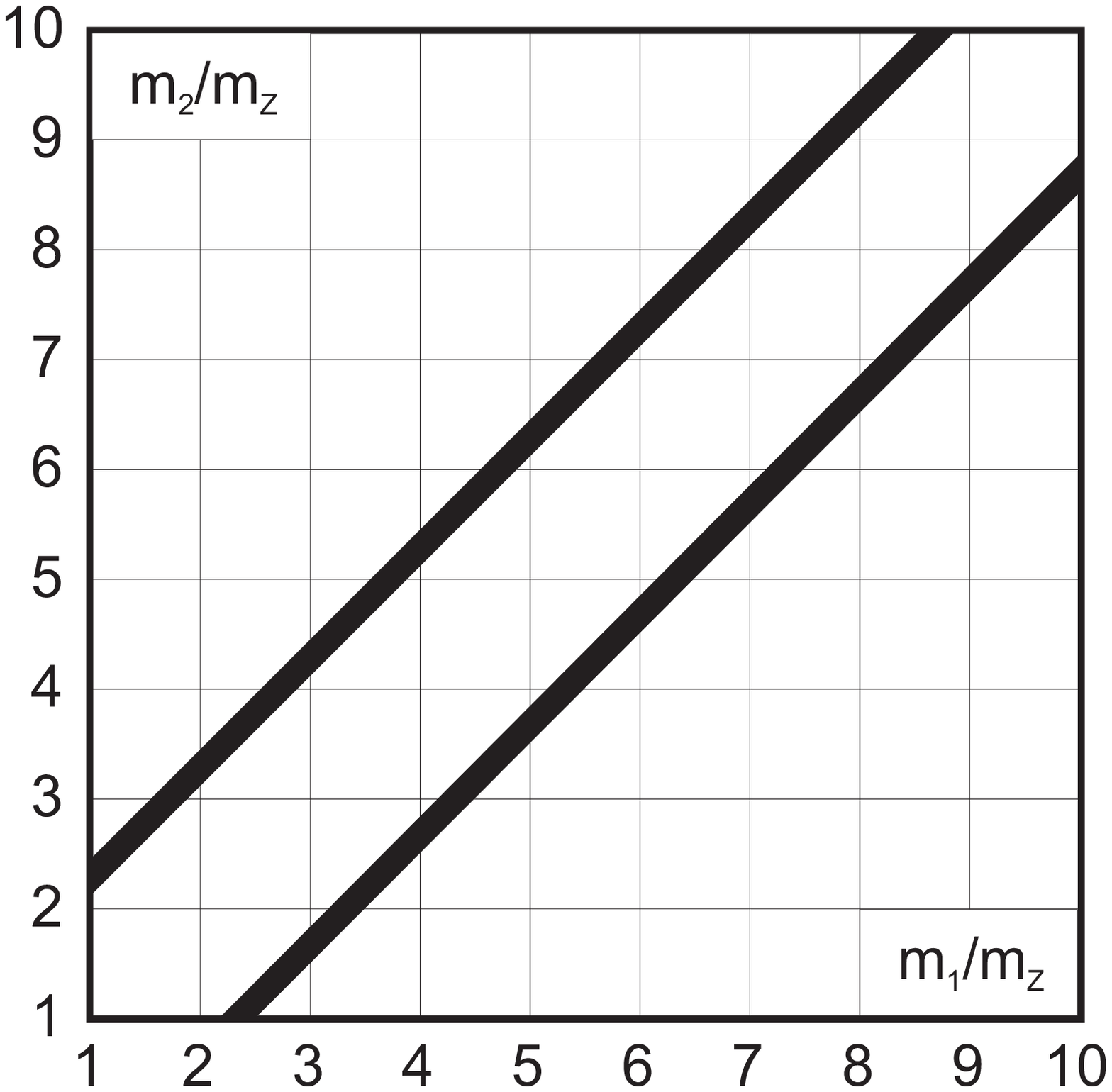}
 }
 \caption{
The shaded areas depict the range for the masses of the new leptons which
are accessible due to the oblique corrections in accordance with the
electroweak precision data without taking into account non-perturbative
corrections. $m_1$ ($m_2$) is the mass, in units of $m_Z$, for the lepton
with the higher (lower)
charge. The black stripes do not correspond exactly to the
overlap of the parabolic area with the 68\% ellipse in the (S,T)-plane from
\cite{unknown:2005em} but with a polygonal area defined by
$-0.1<S+T<+0.5$,
$-0.15<S-T<+0.025$,
and
$S<0.22$.
After taking into account non-perturbative corrections subfigures (b) and
(c)
stay qualitatively the same, while for not too small masses (a) has a second
branch with $m_1<m_2$ like in (c). This corresponds to the overlap of the
ellipse with the right branch of the parabolic area in Fig.~\ref{smleptons}b
as opposed to Fig.~\ref{smleptons}a.
}
\label{masses}
\end{figure*}

Even though it can be considered as a conservative estimate, already the
perturbative assessment of the oblique parameters in our theories shows a
considerable overlap with the data
(see Figs.~\ref{smleptons}a and \ref{frleptons}a). In nearly conformal
theories like ours the contribution of the techniquarks is further reduced by
non-perturbative effects \cite{nonpert1,nonpert2}. This reduction is of the
order of 20$\%$ \cite{nonpert2}. In the case of the integerly charged
leptons (III) the nonperturbative contributions do not change the
characterstics of the results (see Fig.~\ref{frleptons}). The same holds
for the fractionally charged leptons (II). No dedicated plot has been
devoted to that case, because it corresponds to a vertical line exactly in
the opening of the area shaded in black in the other plots. Put differently, 
the black area is contracted to zero width in the direction of $S$. The
situation is slightly different for the standard-model-like charges, where
an additional overlap with the right branch of the black area is
achieved. This corresponds to a second branch in the relative plot shown in
Fig.~\ref{masses}. For our model, the expected mass of the composite Higgs is
150GeV \cite{Dietrich:2005jn}. Let it be noted that, even if it was as heavy
as 1TeV there would still be an overlap between the measurements and the
values attainable in our model.

Translating the overlap depicted in the perturbative versions of
Figs.~\ref{smleptons} and \ref{frleptons} to values of the lepton masses
favoured at the 68$\%$ level of confidence leads to the plots in
Fig.~\ref{masses}. For technical reasons not the exact intersection of the
parabolic shape with the interior of the ellipse is presented but with the
interior of a polygon characterised by:
$-0.1<S+T<+0.5$,
$-0.15<S-T<+0.025$,
and
$S<0.22$. 
In all investigated cases there exists
a branch for which the more negatively charged lepton ($m_2$) is about one Z-boson
mass ($m_Z$) heavier than the more positively charged lepton ($m_1$). The mass gap
of approximately one $m_Z$ is mostly dictated by the limits in the
($S-T$)-direction. The second branch with $m_1>m_2$ is usually forbidden by
the limits imposed on $S$. This does not affect the situation for the
fractionally charged leptons (II), which yield no variation in $S$ as a
function of their masses. Incorporating non-perturbative corrections leads
to a second branch for not too small masses in the standard-model-like
situation (I). This corresponds to the overlap of the ellipse with the right
half of the black area in Fig.~\ref{smleptons}b.


\section{Summary}

In light of the fact that new relevant electroweak precision data have
appeared very recently we have 
investigated  the consequences for the technicolour theory with two 
techniflavours in the
two-index symmetric representation of SU$_\mathrm{T}(2)$ and one additional
lepton generation presented in \cite{Dietrich:2005jn}.
We found that the range of masses of the leptons, consistent 
with the new data at the 68$\%$ level of confidence \cite{unknown:2005em}, is 
much larger than with the previous data at the 90$\%$ level of confidence 
\cite{Eidelman:2004wy}. 
The comparison of our theory with the new precision measurements further 
strengthens our claim that certain technicolour theories are directly 
compatible with precision measurements.


\section*{Acknowledgments}

We would like to thank S.~Bolognesi, S.~B.~Gudnason, C.~Kouvaris, and 
K.~Petrov for discussions.
The work of F.S. is supported by the Marie Curie Excellence Grant under
contract MEXT-CT-2004-013510 and by a Skou Fellowship of the Danish
Research Agency.

\newpage



\section*{Figures}

\begin{thebibliography}{99}

\bibitem{Dietrich:2005jn}
  D.~D.~Dietrich, F.~Sannino and K.~Tuominen,
  Phys.\ Rev.\ D {\bf 72}, 055001 (2005)
  [arXiv:hep-ph/0505059].


\bibitem{TC}
  L.~Susskind,
  Phys.\ Rev.\ D {\bf 20} (1979) 2619;
  S.~Weinberg,
  Phys.\ Rev.\ D {\bf 19} (1979) 1277.


\bibitem{TCrev}
  for recent reviews see:
  C.~T.~Hill and E.~H.~Simmons,
  Phys.\ Rept.\  {\bf 381} (2003) 235
  [Erratum-ibid.\  {\bf 390} (2004) 553]
  [arXiv:hep-ph/0203079];
  K.~Lane,
  arXiv:hep-ph/0202255.


\bibitem{higher}
K.~D.~Lane and E.~Eichten,
Phys.\ Lett.\ B {\bf 222}, 274 (1989);
E.~Eichten and K.~D.~Lane,
Phys.\ Lett.\ B {\bf 90}, 125 (1980);
E.~Corrigan and P.~Ramond,
Phys.\ Lett.\ B {\bf 87}, 73 (1979).

\bibitem{Eidelman:2004wy}
  S.~Eidelman {\it et al.}  [Particle Data Group],
  Phys.\ Lett.\ B {\bf 592}, 1 (2004).


\bibitem{Sannino:2004qp}
F.~Sannino and K.~Tuominen,
Phys.\ Rev.\ D {\bf 71}, 051901 (2005).
arXiv:hep-ph/0405209.


\bibitem{Hong:2004td}
  D.~K.~Hong, S.~D.~H.~Hsu and F.~Sannino,
  Phys.\ Lett.\ B {\bf 597}, 89 (2004)
  [arXiv:hep-ph/0406200].


\bibitem{Cohen:1988sq}
  A.~G.~Cohen and H.~Georgi,
  Nucl.\ Phys.\ B {\bf 314}, 7 (1989).


\bibitem{etc}
  T.~Appelquist and R.~Shrock,
  Phys.\ Lett.\ B {\bf 548} (2002) 204;
  Phys.\ Rev.\ Lett.\  {\bf 90} (2003) 201801;
  N.~D.~Christensen and R.~Shrock,
  arXiv:hep-ph/0509109.


\bibitem{sigma}
 F.~Sannino and J.~Schechter,
 Phys.\ Rev.\ D {\bf 52}, 96 (1995)
 [arXiv:hep-ph/9501417];
 M.~Harada, F.~Sannino and J.~Schechter,
 Phys.\ Rev.\ D {\bf 69}, 034005 (2004)
 [arXiv:hep-ph/0309206];
 D.~Black, A.~H.~Fariborz, F.~Sannino and J.~Schechter,
 Phys.\ Rev.\ D {\bf 59}, 074026 (1999)
 [arXiv:hep-ph/9808415];
 D.~Black, A.~H.~Fariborz, F.~Sannino and J.~Schechter,
 Phys.\ Rev.\ D {\bf 58}, 054012 (1998)
 [arXiv:hep-ph/9804273].


\bibitem{Witten:fp}
E.~Witten,
Phys.\ Lett.\ B {\bf 117}, 324 (1982).


\bibitem{unknown:2005em}
    [ALEPH Collaboration],
  arXiv:hep-ex/0509008.


\bibitem{Sannino:2005dy}
  F.~Sannino,
  arXiv:hep-ph/0506205.


\bibitem{Mahbubani:2005pt}
  R.~Mahbubani and L.~Senatore,
  arXiv:hep-ph/0510064.


\bibitem{Peskin:1990zt}
  M.~E.~Peskin and T.~Takeuchi,
  Phys.\ Rev.\ Lett.\  {\bf 65}, 964 (1990).


\bibitem{nonpert1}
  T.~Appelquist and F.~Sannino,
  Phys.\ Rev.\ D {\bf 59} (1999) 067702;
  T.~Appelquist, P.~S.~Rodrigues da Silva, and F.~Sannino,
  Phys.\ Rev.\ D {\bf 60} (1999) 116007;
  Z.~y.~Duan, P.~S.~Rodrigues da Silva, and F.~Sannino,
  Nucl.\ Phys.\ B {\bf 592} (2001) 371.

\bibitem{nonpert2}
  R.~Sundrum and S.~D.~H.~Hsu,
  Nucl.\ Phys.\ B {\bf 391} (1993) 127;
  M.~Harada, M.~Kurachi and K.~Yamawaki,
  arXiv:hep-ph/0509193.


\end{thebibliography}
\end{document}